\documentclass[twocolumn,american,aps,showpacs]{revtex4}
\usepackage[latin9]{inputenc}
\usepackage{color}
\usepackage{textcomp}
\usepackage{amsmath}
\usepackage{amssymb}
\usepackage{graphicx}
\usepackage{esint}

\makeatletter
\usepackage{dcolumn}
\usepackage{bm}
\usepackage{wrapfig}
\usepackage{subfigure}
\usepackage{ucs}
\usepackage{lineno}
\usepackage{epsfig}

\usepackage{babel}

\makeatother

\usepackage{babel}
\begin{document}

\title{A coupled spin-electron diamond chain with different Land\'e g-factors of localized Ising spins and mobile electrons}

\author{Jordana Torrico$^{1}$, Maria Socorro Seixas Pereira$^{1}$, Jozef Stre\v{c}ka$^{2}$, Marcelo Leite Lyra$^{1}$ }

\affiliation{$^{1}$Instituto de F\'isica, Universidade Federal de Alagoas, 57072-970 Macei\'o, Alagoas, Brazil}

\affiliation{$^{2}$Department od Theoretical Physics and Astrophysics, Faculty of Science, P. J. \v{S}af\'arik University, Park Angelinum 9, 040 01 Ko\v{s}ice, Slovakia}

\begin{abstract}
A coupled spin-electron diamond chain with localized Ising spins placed on its nodal sites and mobile electrons delocalized over interstitial sites is explored in a magnetic field taking into account the difference between Land\'e g-factors of the localized spins and mobile electrons. The ground-state phase diagram is constituted by two classical ferrimagnetic phases, the quantum unsaturated paramagnetic phase and the saturated paramagnetic phase. Both classical ferrimagnetic phases as well as the unsaturated paramagnetic phase are reflected in a low-temperature magnetization curve as intermediate magnetization plateaus. The unsaturated paramagnetic phase is quantum in its character as evidenced by the fermionic concurrence calculated for a pair of the mobile electrons hopping in between the interstitial sites. It is shown that the magnetic field can under certain conditions induce a quantum entanglement above the disentangled ground state.
\end{abstract}


\maketitle

\section{Introduction}
Frustrated spin systems exhibit a variety of exotic quantum ground states, which may provide an interesting alternative for a quantum information processing \cite{nielsen}. The geometric spin frustration is most commonly introduced through competing antiferromagnetic interactions between the localized spins situated on non-bipartite lattices. Another remarkable alternative represents a kinetically-driven spin frustration of the localized spins, which is invoked by a quantum-mechanical hopping process of the mobile electrons. The latter type of spin frustration has been found for instance in the coupled spin-electron diamond chain \cite{pereira1,pereira2,lisnii,torrico1}. Last but not least, recent studies motivated by a magnetic behavior of the copper-iridium oxides have revealed another peculiar mechanism of the spin frustration, which originates from a non-uniformity of the Land\'e g-factor \cite{fireice}.

In the present work, we will explore a combined effect of the kinetically-driven spin frustration and the spin frustration stemming from the non-uniformity of the Land\'e g-factors by generalizing the exact solution for the coupled spin-electron diamond chain \cite{pereira1,pereira2,lisnii,torrico1}. Following the procedure elaborated in our previous work \cite{torrico1} we will compute the fermionic concurrence between the pair of mobile electrons  in order to demonstrate how the magnetic field may induce a quantum entanglement above the disentangled zero-field ground state.

\section{Model and method}
Let us consider a coupled spin-electron diamond chain, which is composed of the localized Ising spins situated on its nodal lattice sites and two mobile electrons hopping on the pairs of interstitial sites (see Fig. 1 of Ref. \cite{torrico1} for illustration). The total Hamiltonian of the model under investigation can be written as a sum of the cell Hamiltonians $\mathcal{H} = \sum_i \mathcal{H}_i$, whereas each cell Hamiltonian $\mathcal{H}_i$ involves all the interaction terms belonging to the $i$-th diamond unit:
\begin{align}
\mathcal{H}_i&=&-t\sum_{\gamma=\uparrow,\downarrow}\left(c_{i1,\gamma}^{\dag}c_{i2,\gamma}
+\mathrm{h.c.}\right)-g_1h\sum_{j=1}^2\left(n_{ij,\uparrow}-n_{ij,\downarrow}\right) \nonumber \\
&+&J\left(\sigma_i+\sigma_{i+1}\right)\sum_{j=1}^2\left(n_{ij,\uparrow}-n_{ij,\downarrow}\right) -g_2\frac{h}{2}\left(\sigma_i+\sigma_{i+1}\right).
\end{align}
Here, $c_{i\alpha,\gamma}^{\dag}$ and $c_{i\alpha,\gamma}$ are fermionic creation and annihilation operators for an electron with the spin $\gamma = \uparrow,\downarrow$ hopping on the pairs of interstitial sites $\alpha=1,2$, $n_{i\alpha,\gamma}=c_{i\alpha,\gamma}^{\dag}c_{i\alpha,\gamma}$ is respective number operator. The parameter $t$ is the hopping term associated with the kinetic energy of the mobile electrons, the coupling constant $J$ stands for the Ising interaction between the nearest-neighbor localized spins and mobile electrons, $h$ is the external magnetic field accounting for the difference between the Land\'e g-factors $g_1$ and $g_2$ of the mobile electrons and localized Ising spins, respectively (Bohr magneton $\mu_{\rm B}$ was absorbed into a definition of the field term $h$).

An exact solution for the coupled spin-electron diamond chain can be straightforwardly obtained by adapting the procedure reported in our previous work \cite{torrico1} by a mere replacement of the uniform magnetic field $h$ through two different local magnetic fields $g_1 h$ and $g_2 h$. The readers interested in details of the calculation procedure are therefore referred to Ref. \cite{torrico1}.

\section{Results and discussion}

Let us proceed to a discussion of the most interesting results for the coupled spin-electron diamond chain with the antiferromagnetic exchange interaction $J>0$, which exhibits the most outstanding magnetic features due to a spin-frustration effect. For simplicity, our further attention will be restricted to the most common special case with the fixed value of the Land\'e g-factor of the mobile electrons $g_1=2$, whereas an influence of the  Land\'e g-factor of the localized Ising spins on the overall magnetic behavior will be subject of our investigations.

\begin{figure}[t]
\includegraphics[scale=0.26]{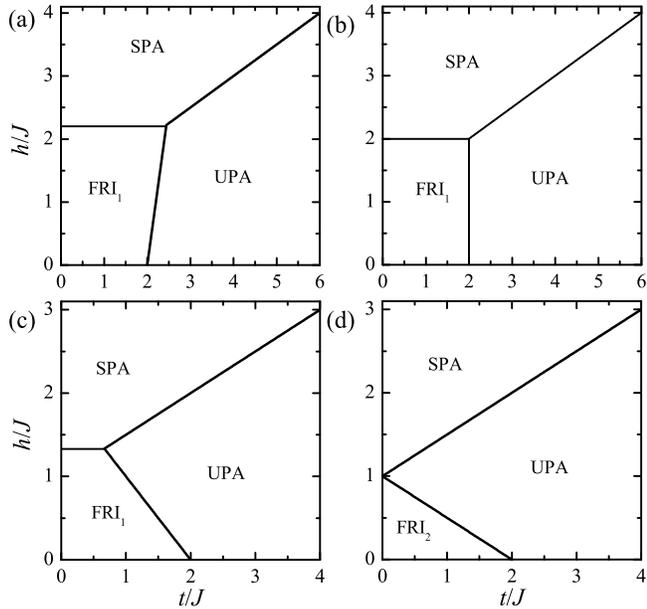}
\vspace*{-0.6cm}
\caption{\label{fig:1} The ground-states phase diagram in the $t/J$-$h/J$ plane for the fixed value of the Land\'e g-factor of the mobile electrons $g_1=2$ and a few different values of the Land\'e g-factor of the localized Ising spins: (a) $g_2=1.8$; (b) $g_2=2$; (c) $g_2=3$; (d) $g_2>4$.}
\end{figure}

\subsection{Ground-state phase diagram}
The ground-state phase diagram (Fig. \ref{fig:1}) involves two ferrimagnetic phases (FRI$_1$ and FRI$_2$), the unsaturated paramagnetic phase (UPA) and the saturated paramagnetic phase (SPA) given by the eigenvectors:
\begin{eqnarray}
\left|\mathrm{FRI}_1\right>&=&\prod_{i=1}^N\left|\downarrow\right>_{\sigma_i}\otimes\left|\uparrow,\uparrow\right>_i \\
\left|\mathrm{FRI}_2\right>&=&\prod_{i=1}^N\left|\uparrow\right>_{\sigma_i}\otimes\left|\downarrow,\downarrow\right>_i \\
\left|\mathrm{UPA}\right>&=&\prod_{i=1}^N\left|\uparrow\right>_{\sigma_i}\otimes \nonumber \\
&~&\frac{1}{2}\left[\left|\uparrow,\downarrow\right>_i+\left|\downarrow,\uparrow\right>_i
-\left|\uparrow\downarrow,0\right>_i-\left|0,\uparrow\downarrow\right>_i\right] \\
\left|\mathrm{SPA}\right>&=&\prod_{i=1}^N\left|\uparrow\right>_{\sigma_i}\otimes\left|\uparrow,\uparrow\right>_i.
\end{eqnarray}
In above, the first ket vector determines the spin state of localized Ising spins and the second one the spin state of the mobile electrons. All magnetic moments of the localized Ising spins and mobile electrons are fully aligned into the magnetic field within the SPA ground state. Owing to the antiferromagnetic interaction $J>0$, the localized Ising spins tend in opposite to the magnetic field within the FRI$_1$ ground state
and the mobile electrons tend in opposite to the magnetic field within the FRI$_2$ ground state. However, the most interesting spin arrangement can be found within the UPA ground state, where the hopping process of two mobile electrons with opposite spins leads to a kinetically-driven spin frustration of the localized Ising spins. As a result, the localized Ising spins are  polarized by arbitrary but non-zero magnetic field, while the mobile electrons underlie a quantum entanglement of two antiferromagnetic and two ionic states. Note furthermore that the FRI$_1$ (FRI$_2$) phase appears in the ground-state phase diagram if $g_2<4$ ($g_2>4$), while both ferrimagnetic phases coexist together when $g_2=4$.

\subsection{Magnetization curves}
\begin{figure}[t]
\includegraphics[scale=0.25]{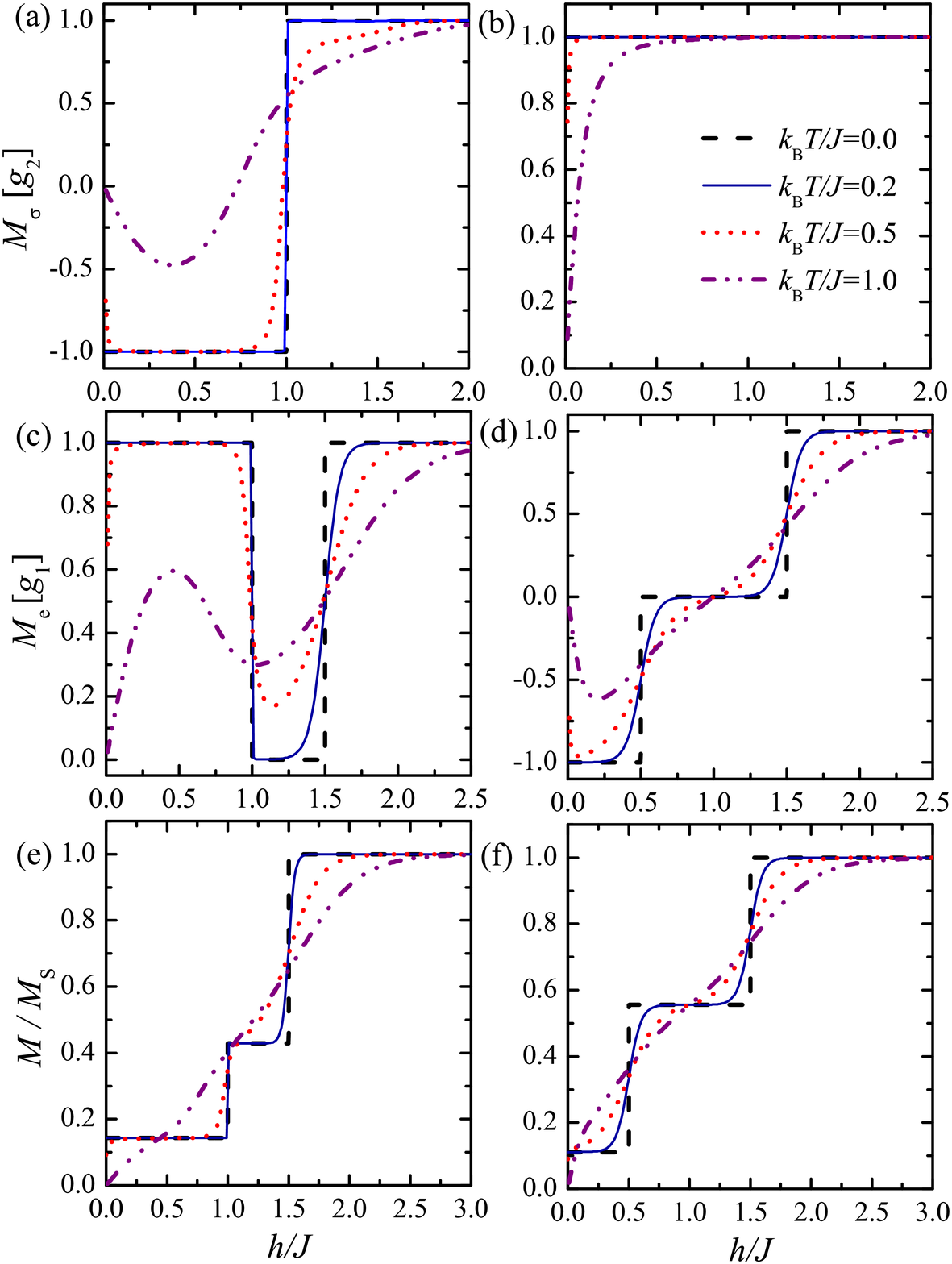}
\vspace*{-0.6cm}
\caption{\label{fig:2} (Color online) The sublattice magnetization of the Ising spins $M_{\sigma}$, the sublattice magnetization of the mobile electrons $M_e$ and the total magnetization as a function of magnetic field for $t/J=1$, $g_1=2$ and a few different values of temperature. The first column corresponds to $g_2=3$ and the second column to $g_2>4$.}
\end{figure}

To get a deeper insight, the total magnetization is plotted in Fig. \ref{fig:2} against the magnetic field together with the sublattice magnetization of the Ising spins and the mobile electrons. The magnetization dependences shown in the first column give evidence for the following sequence of the phase transitions FRI$_1$-UPA-SPA driven by the rising magnetic field. Contrary to this, the total and sublattice magnetizations plotted in the second column give proof for another sequence of the field-induced phase transitions FRI$_2$-UPA-SPA. The displayed magnetization curves thus independently verify correctness of the established ground-state phase diagram.

\subsection{Fermionic concurrence}

\begin{figure}[t]
\vspace*{0.8cm}
\includegraphics[scale=0.25]{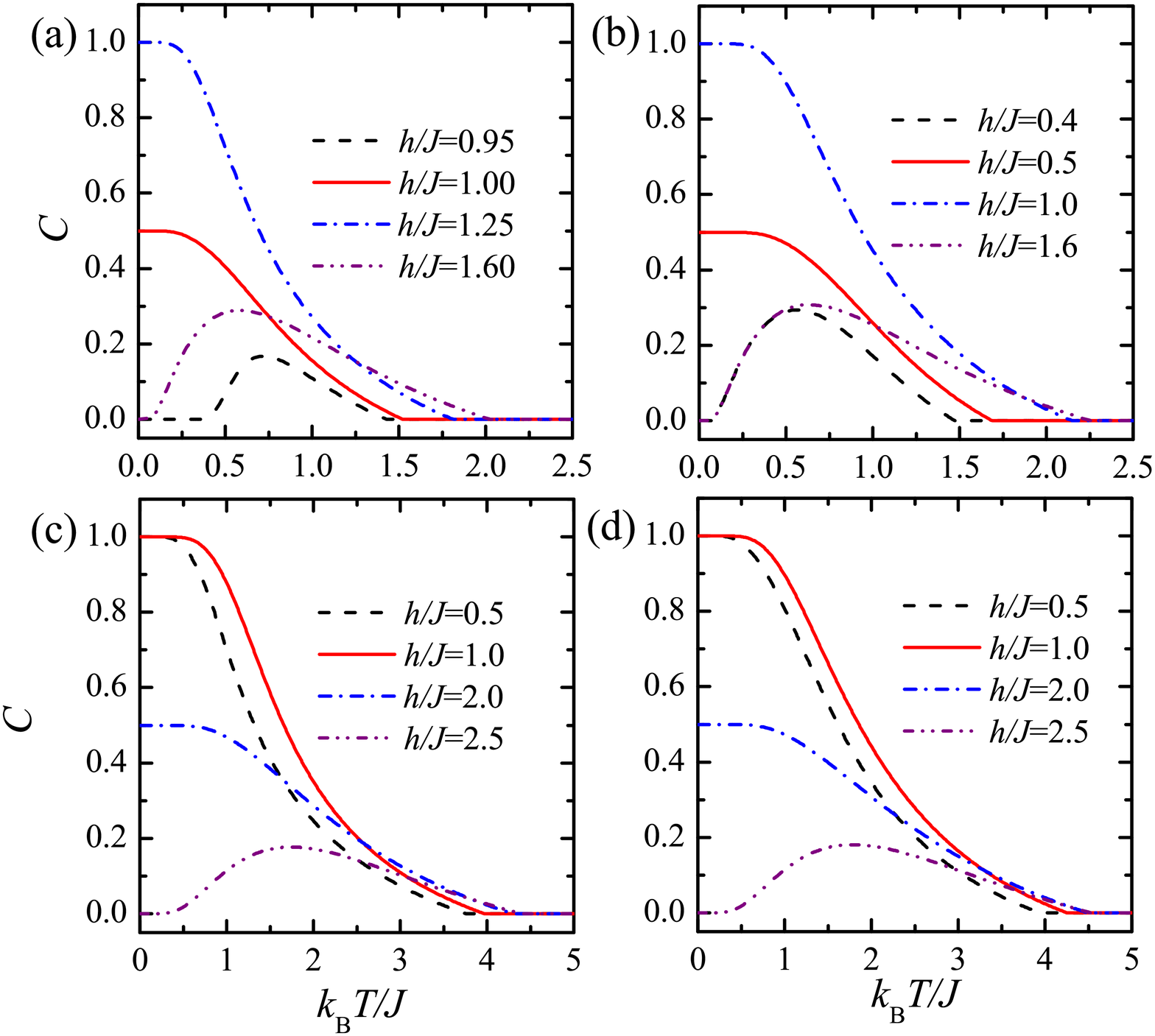}
\vspace*{-0.6cm}
\caption{\label{fig:3} (Color online) Thermal variations of the concurrence at a few different magnetic fields when $g_1=2$ is fixed and:
(a) $t/J=1$, $g_2=3$; (b) $t/J=1$, $g_2>4$; (c) $t/J=2$, $g_2=3$; (d) $t/J=2$, $g_2>4$.}
\end{figure}

The fermionic concurrence $C$, which may serve as a measure of bipartite entanglement between two mobile electrons from the same couple of interstitial sites, can be calculated at zero as well as nonzero temperatures according to the procedure described in Ref. \cite{torrico1}. The classical character of the FRI$_1$, FRI$_2$ and SPA ground states is consistent with zero concurrence $C=0$, while its maximum value $C=1$ reveals a full quantum entanglement  within the UPA ground state. Typical thermal dependences of the concurrence are illustrated in Fig. \ref{fig:3}. In general, the concurrence monotonically decreases from its maximum value $C=1$ with increasing temperature when the interaction parameters drive the investigated system
towards the UPA ground state. In addition, the concurrence exhibits a more striking non-monotonous thermal dependence if the magnetic field
drives the investigated system sufficiently close to a phase boundary between the UPA ground state and one of three classical (FRI$_1$, FRI$_2$ and SPA) ground states. Under this condition, the concurrence shows a peculiar reentrant behavior due to a thermally-induced activation of the UPA spin arrangement, which represents low-lying excited state above one of three classical (FRI$_1$, FRI$_2$ and SPA) ground states.

\section{Conclusion}

In the present work, we have generalized an exact solution for a coupled spin-electron diamond chain by accounting for a difference between the Land\'e g-factors of the localized Ising spins and mobile electrons. The ground-state phase diagram, magnetization process and fermionic concurrence have been investigated in particular. It has been verified that the ground-state phase diagram involves two classical ferrimagnetic phases FRI$_1$ and FRI$_2$, the quantum unsaturated paramagnetic phase UPA as well as the saturated paramagnetic phase SPA. The ground states FRI$_1$, FRI$_2$ and UPA manifest themselves in a low-temperature magnetization curve as intermediate magnetization plateaus. The quantum character of the UPA ground state has been evidenced through the fermionic concurrence, which displays monotonous decline upon rising temperature. It has been also demonstrated that the nonzero fermionic concurrence can be induced above the classical ground states once the magnetic field drives the investigated system sufficiently close to a phase boundary with the UPA ground state.

\section*{Acknowledgments}

This work was supported by Brazilian agencies FAPEAL, CNPq, CAPES and by Slovak Research and Development Agency under contract No. APVV-0097-12.

\end{document}